\documentclass[11pt,a4paper]{article}
\usepackage{color,amsmath,amssymb,soul,graphicx} 
\usepackage{authblk}
\usepackage[square, numbers, sort&compress]{natbib}

\usepackage[left=45mm,top=25mm,right=35mm,bottom=40mm]{geometry}
\title{Long-range synchrony and emergence of reentry \protect\\   in  neural networks}
\author{Hanna Keren \& Shimon Marom}
\affil{\textit{Network Biology Research Laboratory, Electrical Engineering Department of Physiology, Biophysics and Systems Biology, Medicine Technion - Israel Institute of Technology}}

\begin{document}
\date{}
\maketitle

\begin{abstract} 

\noindent Neural synchronization across long distances is a functionally important phenomenon. In order to  access the mechanistic basis of long-range synchrony, we constructed an experimental model that enables monitoring of spiking activities over centimeter scale in networks of cortical neurons. We show that the mode of synchrony over these distances depends upon a length scale, $\lambda$, which is the minimal path that activity should travel through before meeting its point of origin ready for reactivation. When $\lambda$ is experimentally made larger than the physical dimension of the network, distant neuronal populations operate synchronously, giving rise to irregularly occurring network-wide events that last hundreds of milliseconds to several seconds. In contrast, when $\lambda$ approaches the dimension of the network, a continuous self-sustained reentry propagation emerges, a regular dynamical mode that is marked by precise spatiotemporal patterns (`synfire chains') that may last many minutes. Termination of the reentry phase is due to decrease of propagation speed to a halt. The results contribute to discussions on the origin of different modes of neural synchrony, in normal and pathological conditions.
\end{abstract}

synchronization | length-scale | reentry | synfire-chains | disinhibition 
\section{Significance Statement} 

The mode of synchrony a neural network resumes is a critical determinant of its function. The number of microscopic mechanisms that impact on the macroscopic mode of synchrony is immense. Here we extract a lumped physical parameter ($\lambda$) and show that its value controls the mode of synchrony in a network of biological neurons. As such it promotes the understanding of synchrony modes otherwise masked by the richness of underlying microscopic complexity. The hope is that the insights gained in this study will cater to manipulation of the phenomena in pathological conditions.

\section{Introduction}

Spontaneous synchronization between remote neural networks is one of the hallmarks of brain activity, considered significant for multiple functions in health and disease \cite{Steriade1994,Rodriguez1999a, Engel2001b,Varela2001,bianchi2012network,Jasper2012}. 
As such, the study of mechanisms underlying long-range synchronized activity is of substantial interest.  In the cortex, long range centimeter scale synchronization involves an interplay between the statistics of synaptic connectivity amongst excitatory neurons, and the global effect of rapidly propagated activity through electrically coupled inhibitory neurons  \cite{ Gibson1999, Beierlein2000,   Sanchez-Vives2000,Linkenkaer-Hansen2001, Volgushev2006}. 
There are many different \textit{microscopic parameters} at the cellular and synaptic levels that impact on the above interplay between excitation and inhibition \cite{Destexhe2003, Destexhe2004, Cline2005, Marder2006g}. But here we are interested in exposing a \textit{global physical parameter} that mediates the translation of the many microscopic mechanisms to macroscopic modes of long-range synchrony. Such parametrization has been useful in analyzing cardiac reentry arrhythmias \cite{Winfree1989}; 
measurements of cyclic brain activity suggest that a similar approach might be instructive for the case of neural synchrony \cite{Huang2010}. To this aim, we take advantage of a reduced experimental model of large-scale networks composed of randomly connected cortical neurons. We show that the mode of synchrony is sensitive to a characteristic length scale, $\lambda$, which is the product of two parameters -- the time scale ($\tau$) of the individual synchronous event (including its refractory period), multiplied by the speed of activity propagation ($\nu$). When this length scale exceeds the longest propagation path offered by the network, the system acts as a single compartment, being active simultaneously. In contrast, when the characteristic length scale is in the range of propagation paths supported by the network, self-sustained activities appear, similar to the familiar reentrant dynamics in cardiac arrhythmia. We manipulate the relevant parameters ($\tau$ and $\nu$) using pharmacological blockade of GABA$_{A}$ mediated inhibition and show that their relations determine the global mode of synchrony: from network-wide simultaneous activity that occurs irregularly, to regular ongoing reentry dynamics that may last many minutes and runs through temporally precise synfire-chains.


\section{Results and Discussion}

\subsection{Long-range synchronization}

When allowed to develop for a couple of weeks outside the brain, a population of cortical neurons (extracted from the rat newborn) tends to form a large-scale network that exhibits complex spontaneous activity. This spontaneous activity is characteristically composed of synchronous events (a.k.a. `network spikes') that occur irregularly at an average rate of ca.~$0.1$ Hz, interspersed with some uncorrelated sporadic activities \cite{509, corner2002, Marom2002d, Morin2005, EytanMarom2006, Wagenaar2006b}.
 The temporal envelope of each synchronous event is usually monitored by integrating the spiking activities detected through many individual electrodes (in the order of several tens) arranged in an array. In most cases the millimeter scale electrode array is positioned at the center of the large-scale network (Figure 1A, left panel). Monitored through such relatively dense array of electrodes at the center, the time scale of a single synchronous event -- a single network spike -- is in the order of one hundred milliseconds, as depicted in Figure 1A (middle and right panels) and discussed elsewhere \cite{EytanMarom2006}.

In order to adjust the above standard experimental design to the subject matter of the present study that involves distances an order of magnitude larger, we constructed an electrode array layout consisting of four clusters that cover a significantly wider area (Figure 1B, left panel; different colors depict the physical identity of each recording region). As can be appreciated by the examples provided in the middle and right panels of Figure 1B, when activity is integrated over such distances, the apparent time scale of a single synchronous event is accordingly extended ($1.02 \pm 0.47$ seconds; n=1721, 8 networks), a value that is comparable to durations observed \textit{in-vivo} \cite{Steriade1993c, Sanchez-Vives2000}. Also note that activity propagates in a wave-like manner across regions, although termination is less ordered (colored arrows). 

\vspace{+3em}
\begin{figure*}[h!]
\begin{center}
\centerline{\includegraphics[width=124mm]{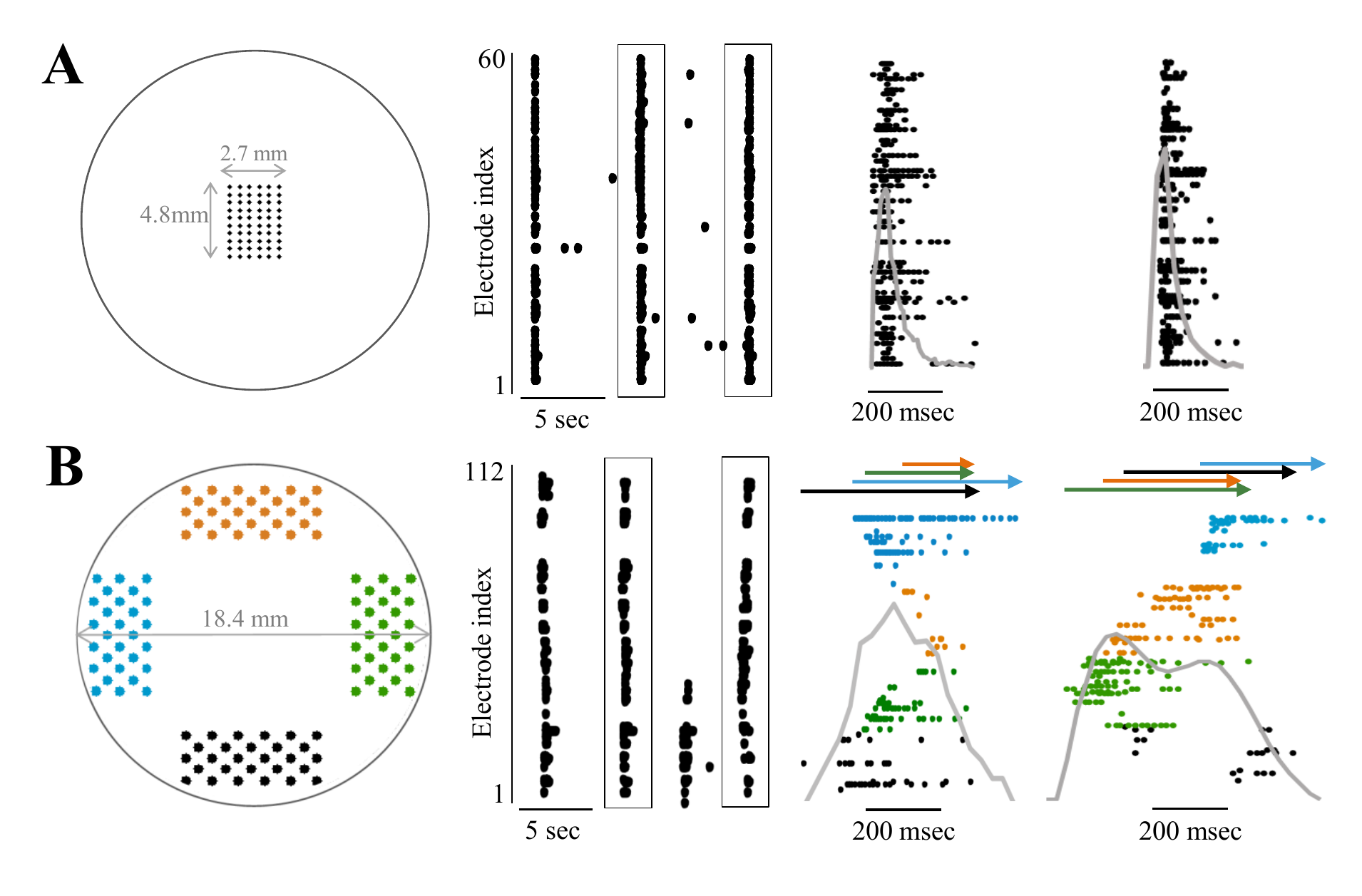}}
\caption{\small{\textbf{Recording long-range synchronization}. (\textbf{A}) Scheme of a standard centered array of 60 electrodes. Three examples of spiking activity monitored through such an array are shown at the middle panel. Each point depicts a single spike detected in one of the electrodes (indexed in the vertical axis). Two synchronous events (boxed) are presented at higher temporal resolution on the right-hand panel. A superposed gray line shows normalized summed activity (binned to 25 msec). (\textbf{B}) Similar presentation of activity recorded using a layout of 112 electrodes arranged in four clusters of recording regions (color coded). Arrows demonstrate the duration of activation in each of the regions in respective colors.}}
\end{center}
\end{figure*}
\vspace{+4em}

The region wherefrom synchronous activity is initiated varies between synchronous events. Figure 2A shows that the path of propagation amongst the four regions is uniquely determined by the region of activity initiation; this is indicated by the recruitment profiles of several synchronous events, color coded according to initiation region. The averaged delay between activities recorded in different regions is 80 msec and is fairly consistent in the eight different networks studied here (Figures 2B-C). The calculated speed of propagation ($\nu$) is 0.18 meter per second (SD=0.08, n=1721 synchronous events), which is within the range of reported propagation speed in the primate cortex \cite{Thiagarajan2010c}. The interval ($\tau$) between two subsequent initiations of network spikes is broadly distributed, spanning a range of several seconds ($\tau = 10.3$ seconds, SD=11, n=1721). The standard deviation of $\tau$, which is in the order of the mean, is mainly contributed by variations in the inter-event-intervals rather than the actual network spike shape. The minimal value of $\tau$, composed of the synchronous event itself and the recovery period that follows, is rarely ($<5\%$) shorter than 3--4 seconds \cite{EytanMarom2006, Wagenaar2006b, Haroush2015}.

Note that the complex nature of connectivity statistics may be appreciated by observing that earlier spiking activity in a given region does not entail a respective earlier peak of the synchronous event within that region. In 17\% of the 1721 network spikes examined, a synchronous event initiated in one region reaches peak population activity in a downstream region earlier than in its upstream originator (inset to Figure 2C). This observation reflects, most likely, a complex graph of connectivity and the presence of hub neurons, as discussed elsewhere \cite{EytanMarom2006, Buzsaki2010, Luccioli2014}. 

The product $\tau\cdot\nu$ yields a characteristic length scale ($\lambda$) that sets a constraint on the possible mode of synchrony. It provides an estimate for the minimal length of path through which activity should travel before meeting its point of origin ready for reactivation. The networks studied here are randomly spread on top of a circular plate having a perimeter of $\approx 6$ centimeters. Therefore, the numbers provided above ($\tau > 3$ sec, $\nu \approx 20$ cm/sec) yield a characteristic length that is more than an order of magnitude larger compared to the dimensions of the system, entailing a single-compartment-like behavior. 

Consistent with the above, the results presented in Figure 2D-F demonstrate simultaneous termination of synchronous activity at distant sites on a centimeter scale. In Figure 2D, four traces of  synchronous events -- averaged from 8 different networks -- are shown, classified according to the \textit{rank-order of activation times}: The leftmost trace (depicted by arrow 1) is the average activity (spikes per msec) recorded in the first-to-synchronize region, regardless of its physical location. Note that, congruent with the data of Figure 2A, different regions contributed to the average of this leftmost trace, as each global synchrony event could have originated in any one of four different physical regions. Likewise, the traces depicted by arrows 2--4 are the average activities recorded in the second, third and fourth-to-synchronize regions, respectively, regardless of their physical identities. On \textit{average}, the envelope of a global synchronous event is dictated by the first-to-synchronize region; synchronized activity in all downstream regions converge to that envelope. This is also manifested as decreased amplitudes and durations of synchronized events in downstream regions (Figures 2E-F). Overall, in spite of consistent offset in the initiation of synchronized activity -- fully accounted for by propagation delays -- on average there is no appreciable offset in the termination of remote activities; the process of synchrony cessation over relatively long distances seems literally simultaneous. 

The phenomenon of simultaneous termination, without which a large-scale network can fall into modes of self-sustained waves, has been hitherto demonstrated in several animals \textit{in-vivo}, using single and multi-unit recordings, as well as EEG, and suggested to occur due to the inhibitory cells being strongly coupled via gap junctions \cite{Volgushev2006, Chen2012d}.  

\begin{figure*}[h!]
\begin{center}
\centerline{\includegraphics[width=118mm]{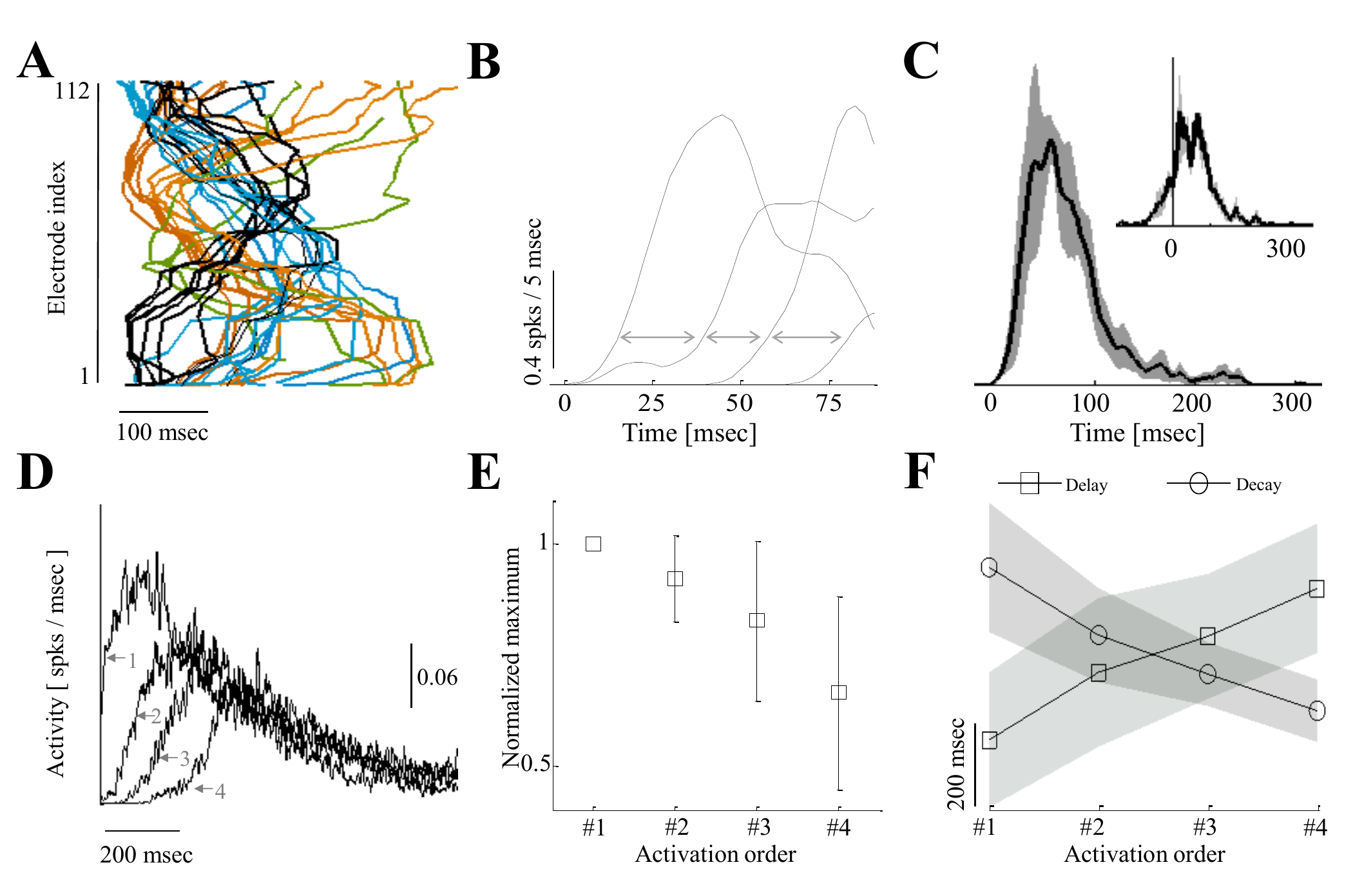}}
\caption{\small{\textbf{Propagation and termination of synchronized activity across remote regions}. (\textbf{A}) Several synchronous events presented according to the temporal order (horizontal axis) of the participating electrodes (vertical axis). Color is indicative of the identity of the recording region where synchrony was initiated. Propagation delays between recording regions are exemplified for a single synchronous event in (\textbf{B}) (gray arrows). The average distribution of such delays between initiation times of temporally adjacent regions is shown in panel \textbf{C} (eight networks, 1721 events; standard deviation across networks depicted by shaded area). Inset: distribution of delays between times of maximal activity of regions, sorted according to the initiation time of each region. Note negative values that represent a mismatch between the order of initiation and the order of arriving at maximal activity. (\textbf{D}) All synchronous events (n=1721, 8 networks), averaged and classified according to the rank-order of initiation times. (\textbf{E}) Values of maximal activity for each region, averaged and ranked according to initiation times. For each event, values of second (\#2) to fourth (\#4) recruited regions are normalized to maximal activity of the first-to-fire region (\#1). Error bars represent standard deviation across all events. (\textbf{F}) Time delay from detection of initiation of activity in a given recording region, to maximal activity in that same region; this value (square symbols) is a proxy for within-region recruitment rate. Circles depict time from maximal activity of a given region to first occurrence of no activity within that same region, indicating within-region decay time. Shaded margins are standard deviation across all events.}}
\end{center}
\vspace{-2em}
\end{figure*}

\newpage

\subsection{Manipulating the characteristic length scale by disinhibition}

We found GABA$_A$ blockers (Bicuculline or Picrotoxin) very useful as means to explore the effects of the product $\tau\cdot\nu$ on modes of synchrony. This is due to the fact of both blockers demonstrating a seemingly paradoxical effect of increasing $\tau\cdot\nu$ at low concentrations, while decreasing it at high concentrations. 
As demonstrated in Figures 3A-B, partial GABA$_{A}$ blockade using ca.~5 $\mu$M Bicuculline (IC50 = 3$\mu$M) significantly prolongs the duration of synchronized activity, from the range of hundreds of milliseconds to several seconds ($3.69 \pm 1.41$ seconds; n=1721, 8 networks). Recruitment toward synchrony becomes faster and propagation speed $\nu$ is increased to $0.43$ meter per second (SD=0.19, n=1143 events from 8 networks). The values of $\tau$ are also increased (average 28 sec, SD=16, n=1143, 8 networks) and, as a consequence, the characteristic length scale becomes very large compared to control condition. The resulting network-wide synchrony mode is marked by extended (several seconds) relaxations. Termination across the network, while less simultaneous in absolute terms, is not much different from control condition when variation is normalized to the mean duration of the network spike. These observations are congruent with previous reports on the effect of GABA$_{A}$ disinhibition in similar concentrations \cite{EytanMarom2006,Chen2012d}.


\begin{figure*}[ht!]
\vspace{-0.5em}
\begin{center}
\centerline{\includegraphics[width=160mm]{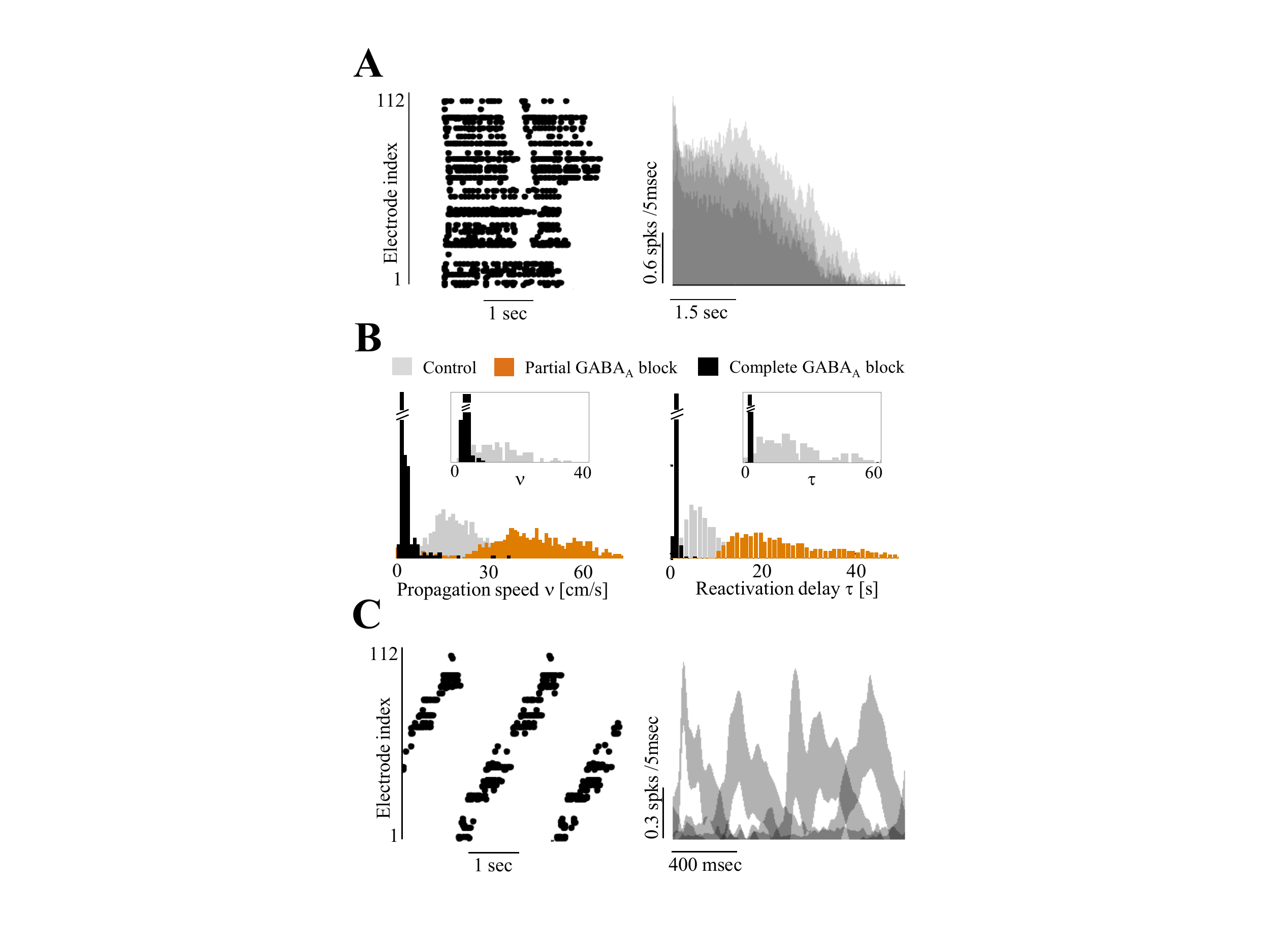}}
\caption{\small{\textbf{Manipulating characteristic length scale by disinhibition}. (\textbf{A}) Left: A single synchronous event under \textit{partial} Bicuculline block (5$\mu$M), each point depicts a single spike detected in one of the electrodes (indexed in the vertical axis). Right: 146 events from a single network under this condition, classified and ranked according to region initiation time; for visual clarity only the positive margins of the standard deviation across all events are presented. (\textbf{B}) Distributions of propagation speed $\nu$ (left) and reactivation delays $\tau$ (right) under three conditions: \textit{control} (no blockers; light gray, n=1721, 8 networks), \textit{partial GABA$_{A}$ block} (5$\mu$M Bicuculline; dark gray, n=1143, 8 networks) and \textit{complete GABA$_{A}$ block} (mean concentration of 100$\mu$M Bicuculline; black, n=478, 8 networks). Propagation speed was estimated from the initiation delays between recording regions; histograms constructed with 1 cm/sec bins. Reactivation delays are the intervals between subsequent synchronous events across the whole network; histograms constructed with bins of 600 msec. Insets are the corresponding analyses of 3 networks with complete Picrotoxin block (200$\mu$M; black, n=347) and control (light gray, n=157). (\textbf{C}) Example of activity under \textit{complete} Bicuculline block. Right: standard deviation margins of 114 events under this condition.}}
\end{center}
\vspace{-2em}
\end{figure*}

Intriguingly, under different settings, other (even opposite) impacts of disinhibition on the duration of synchronization were reported  \cite{Mann2009, Sanchez-Vives2010}. Maybe related, Chen et al. \cite{Chen2012d} suggested, based on numeric simulations, that further elimination of inhibition might give rise to an opposite effect of \textit{reduced} duration of synchronous events, due to fast activation of hyperpolarizing conductances. With this in mind, we attempted pushing the product $\tau\cdot\nu$ to values lower compared to control condition. Indeed, we observe a non-monotonous effect of Bicuculline as concentration of the blocker is further increased. Stronger GABA$_{A}$ block by Bicuculline slows down the propagation speed ($\nu$) to 0.042 meter per sec (SD=0.015, n=478, 8 networks); this is reflected in longer delays between activities of adjacent regions (Figure 3B, left). In addition, at these higher concentrations, the blocker decreases the average $\tau$ to 1.52 seconds (SD=0.44, n=478, 8 networks); this is due to shortening of within-region synchronous event duration, as well as shortening of the inter synchronous intervals (Figure 3B, right). We note a marked variability in the concentration of Bicuculline required to elicit these effects in the four networks tested here, the average being 100 $\mu$M. The mechanism underlying the above non-monotonous effects of Bicuculine is not clear, and published reports cast doubt on whether at all they are related to inhibitory synaptic transmission. Specifically, Bicuculline (methiodide derivative) has been shown to impact on membrane ionic conductances by blocking calcium-activated potassium channels \cite{Khawaled1999}. We therefore replicated the main observations related to high Bicuculline concentration using Picrotoxin -- a different class of GABA$_{A}$ block \cite{ Khawaled1999, Johansson2001}
 -- using a concentration indicated to induce complete blockade of GABA$_{A}$ inhibition (200 $\mu$M \cite{Wang1995}). Insets of Figures 3B show that the effect of Picrotoxin on both propagation speed $\nu$ and reactivation delays $\tau$ is similar to that of high concentrations of Bicuculline, suggestive of a possible mechanism that does involve inhibition. Regardless of the machinery underlying the effect of high Bicuculline or Picrotoxin concentrations, from the perspective of the present report what matters is that the synchronization mode under these conditions is characterized by slowly propagated activity that runs through adjacent regions (Figure 3C). 

\subsection{Reentry mode}
More specifically, complete GABA$_{A}$ block reduces the product $\tau\cdot\nu$ from ca.~200 cm in control conditions to a value close to the physical perimeter of the network (slightly larger than 6 cm). As a consequence, emergence of reentrant -- self sustained (reverberating) patterns -- becomes probable, and Figure 4 demonstrates several realizations of such a mode. Phases of reentry may last many minutes (Figure 4A), and the direction of the reentrant activity may flip between clockwise to counter-clockwise (Figure 4B). The preciseness of the spatial and temporal patterns emerging under this condition is demonstrated in Figure 4C using terms borrowed from Schrader \textit{et al.} \cite{Schrader2008} in their analyses aimed at identification of synfire-chains \cite{Abeles1991, Abeles1993}. 

\begin{figure*}[h!]
\begin{center}
\centerline{\includegraphics[width=129mm]{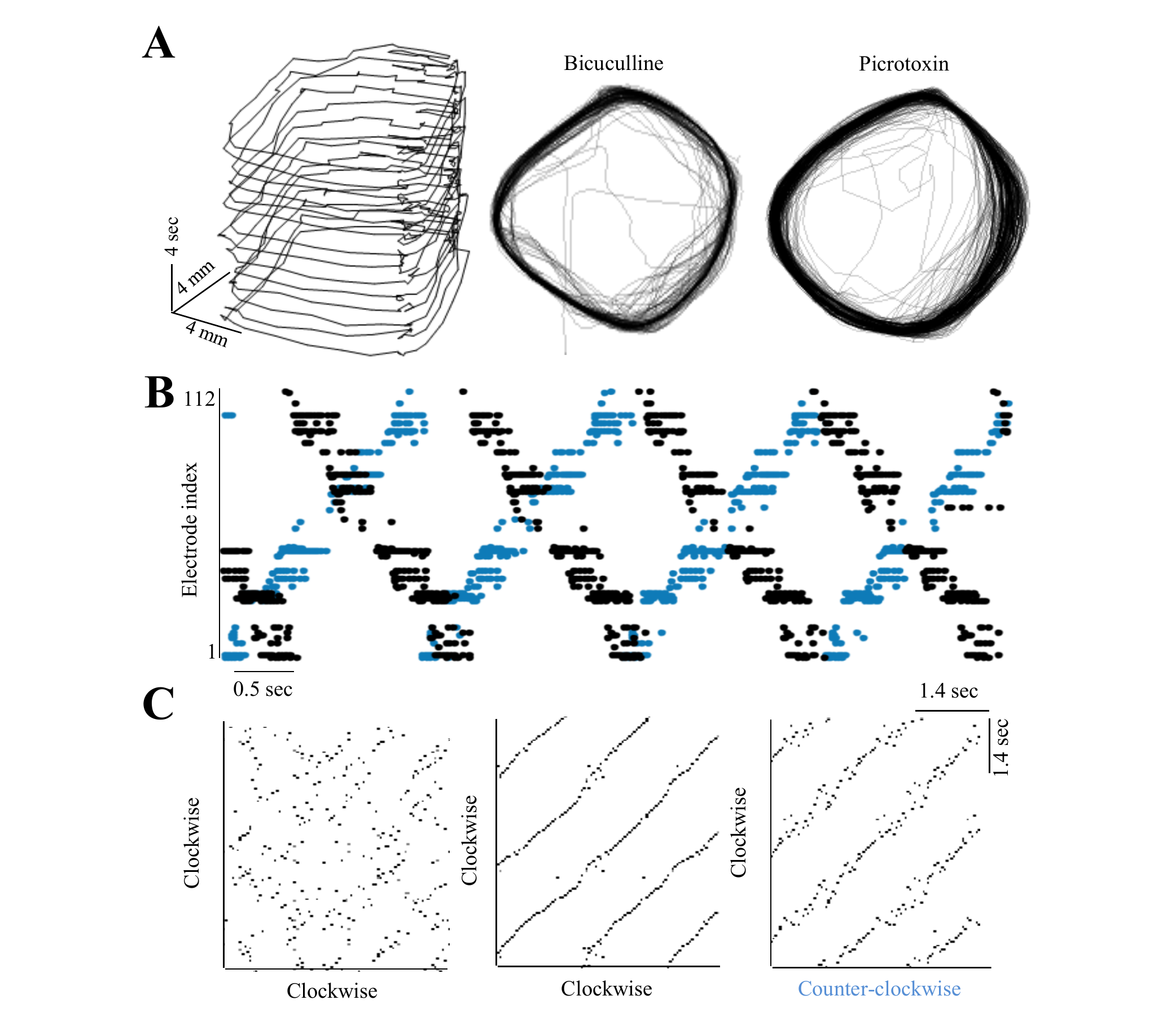}}
\caption{\small{\textbf{Reentry in disinhibited networks}. (\textbf{A}) Coordinates of average activity position during reentry, calculated from the first spike detected by each electrode per cycle (moving average over 5 recent spikes). The right hand panels present two entire phases of reentry in other networks under complete Bicuculline or Picrotoxin block, lasting 3.3 and 8.3 min. (\textbf{B}) Two frames extracted from two consecutive phases of reentry in the same network; the phases run in opposite spatial directions (color coded). (\textbf{C}) Synfire chains detection procedure using a pair-wise similarity matrix following Schrader et al.~\protect\cite{Schrader2008}: Each pixel represents the level of similarity (scaled 0--1) between sets of active electrodes in two time bins (3 msec). Black depicts full identity (see Methods). The left-hand panel is calculated under control condition. The middle panel shows synfire chains detected in a clock-wise directed reentry activity. The right-side of panel of \textbf{C} is a comparison between a reversed \textit{counter clock-wise} reentry propagation and a clock-wise data.}}
\end{center}
\vspace{-3em}
\end{figure*}

Intriguingly, comparison of activity patterns of clockwise and counter-clockwise reverberations (in a given network) suggests that waves of activity travel through (more-or-less) the same path, but in opposite directions (Figure 4C right panel). This implies that under complete disinhibition, the spatial correlation length is shorter than the ca.~1 mm that separates nearby electrodes within a region.

Termination of a reentry phase occurs abruptly, as shown in Figure 5A. As proposed by Golomb and Amitai \cite{Golomb1997}, network activity comes to a halt once propagation speed is below 5 cm/sec, due to reduced excitatory synapses efficacy (of type AMPA). Indeed, in our case (Figure 5B) termination is preceded by a gradual decrease of propagation speed until a threshold value of ca.~3 cm/sec is reached (mean=3.42, SD=0.26, across 44 reentry phases in 6  different networks), as well as a gradual decrease in population recruitment rate (Figure 5C).


\begin{figure*}[h!]
\vspace{+2em}
\begin{center}
\centerline{\includegraphics[width=80mm]{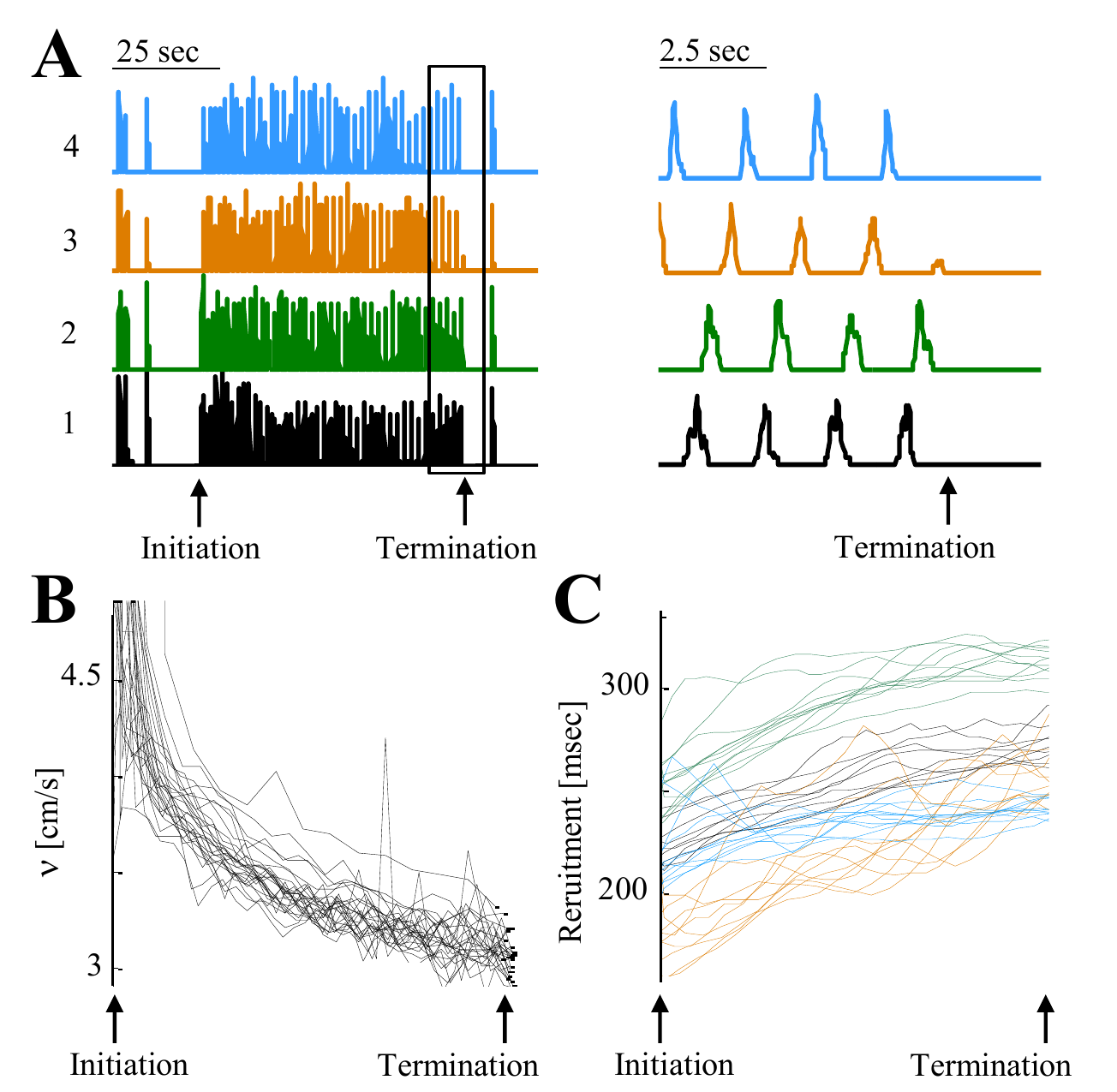}}
\caption{\small{\textbf{Termination of reentry phases}. (\textbf{A}) The case of one reentry phase; the termination phase (boxed) is shown on the right panel at higher temporal resolution. (\textbf{B}) Decline of propagation speed in 27 reentry phases of the same network, aligned according to initiation and termination times of each reentry phase. (\textbf{C}) Similar presentation of recruitment duration, estimated for each synchronous event from the time elapsed until all participating electrodes of a given region (color coded) detected at least one spike.}}
\end{center}
\end{figure*}

And, finally, in a case where reentry waves run along the perimeter of a circular network, activity at the center area is expected to be less stereotypic. Indeed, when reentry commences, the activity recorded from several electrodes located at the center of the array is significantly reduced, and becomes less correlated with the activity picked by the peripheral electrodes (Figure 6).

\begin{figure*}[h!]
\vspace{-3.5em}
\begin{center}
\centerline{\includegraphics[width=125mm]{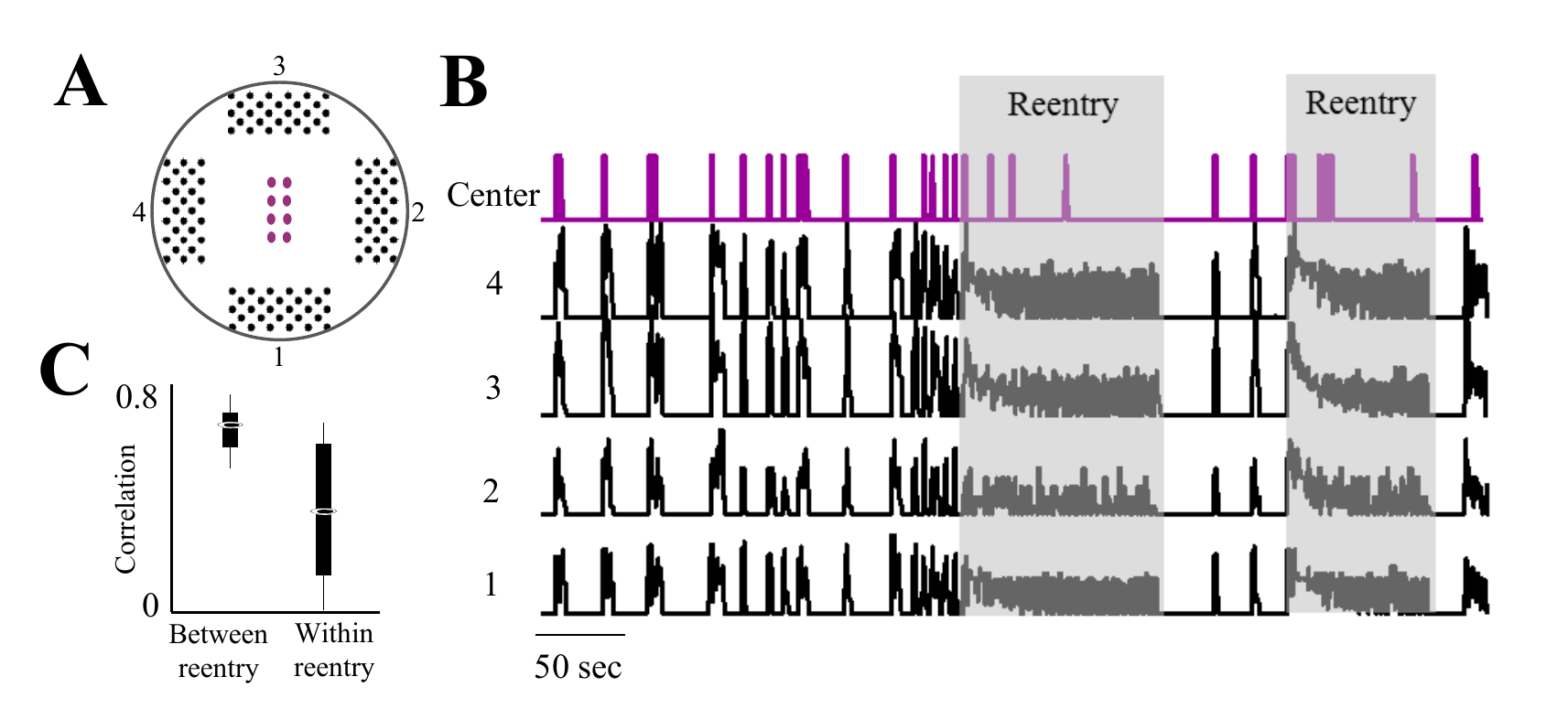}}
\caption{\small{\textbf{Activity detected in the central area during reentry propagation}. (\textbf{A}) A scheme of a recording array that includes 8 electrodes in the central region. (\textbf{B}) Exemplar activities detected in the four peripheral regions (black) and one electrode located in the center (purple), under complete Bicuculline block. Activity was binned to 250 msec. In between reentry phases, the activity detected in central electrodes is highly correlated with activity in peripheral electrodes; within reentry, this correlation drops. This is shown in panel (\textbf{C}), summarizing results from 11 central electrodes in 6 different networks. }}
\end{center}
\vspace{-3.5em}
\end{figure*}    

\newpage
\textbf{Concluding remarks:} \textbf{(1)} By implementing a centimeter scale \textit{in-vitro} experimental design and manipulating $GABA_{A}$ mediated inhibitory activity, we demonstrate that a characteristic length scale, $\lambda$ -- the product of the time scale ($\tau$) of a single synchronous event (including refractoriness) and the speed of activity propagation ($\nu$) -- determines the mode of synchrony between distant network areas. When the longest propagation path in a random network, dictated by the network dimensions, is in the range of $\lambda$, the network tends to fall into a self-sustained reentrant mode of synchronous activity.  Termination of a reentry phase is preceded by a gradual decrease of propagation speed to a lower boundary. We further show that the length scale is sensitive, in a non-trivial manner, to the level of inhibitory transmission. \textbf{(2)} Sampling at a relevant spatial scale matters, as interpretation of underlying mechanism might be biased when monitored through a one given region. For instance, we and others insisted, for many years now, to record the activity of large-scale random networks through a cluster of electrodes positioned in the center of the network, thus avoiding boundary effects. Given the results reported here, focusing on activity at the center of a disinhibited network might mask the coherent nature of ongoing well-structured reentry dynamics, that can only be captured by spreading recording electrodes over the appropriate scale. Likewise, when focusing on one peripheral region we might have interpreted disinhibition-induced reentry as a focal source of short synchrony events, matching commonly observed epileptic-form activity \cite{Chervin1988, Sanchez-Vives2000, Chen2012d}. And, \textbf{(3)} The results of our experimental analyses are not confined to the actual physical structure used here (circular arrangement); what matters is the ratio between the characteristic length and the longest propagation path supported by the network. Indeed, spontaneous occurrence of repeated cyclic propagation has been detected also \textit{in-vivo}, in the absence of clear physical boundary conditions \cite{Huang2010}. Therefore, the message conveyed might be relevant to discussions on the origin of reverberating activity in normal conditions \cite{Edelman1993, Diesmann1999, Durstewitz2000, Boehler2008,  Barak2014, Seelig2015}, 
and forms of epileptic seizures, described both \textit{in-vivo} as well as \textit{in-vitro} \cite{Prince1978, Steriade1998a, Treiman2001a, GarciaDominguez2005,Sanchez-Vives2010, Jasper2012}. 
\vspace{-3.5em}

\vspace{-3em}

\newpage

\section{Material \& Methods}
\subsection{Cell preparation} Cortical neurons were obtained from newborn rats (Sprague-Dawley) within 24 hours after birth using mechanical and enzymatic procedures described in earlier studies \cite{Marom2002d}. Rats were anesthetized by CO$_{2}$ inhalation according to protocols approved by the Technion's ethics committee. The neurons were plated directly onto substrate-integrated multi electrode arrays and allowed to develop into functionally and structurally mature networks over a period of 2-3 weeks. The number of plated neurons was of the order of 450,000, covering an area of 380 mm$^2$. The preparations were bathed in MEM supplemented with heat-inactivated horse serum (5\%), glutamine (0.5 mM), glucose (20 mM), and gentamycin (10 $\mu$g/ml), and maintained in an atmosphere of 37$^{\circ}$C, 5\% CO$_{2}$ and 95\% air in an incubator as well as during the recording phases.
Inhibitory synaptic transmission was blocked by acute application to the bathing solution of either Bicuculline-Methiodide or Picrotoxin (Sigma-Aldrich).

\subsection{Electrophysiology} An array of 112 TiN extracellular electrodes of 30 $\mu$m in diameter, was used (MultiChannelSystems, Reutlingen, Germany). In this array the nearest electrodes are spaced 990 $\mu$m from each other. Recording of central region activity was performed using two columns of 4 electrodes separated by 800 $\mu$m, located in the center of the culture. The insulation layer (silicon nitride) was pre-treated with polyethyleneimine (Sigma, 0.01\% in 0.1 M Borate buffer solution). A commercial amplifier (MEA2100, MCS, Reutlingen, Germany) was used, analog data was low-pass filtered (9 KHz) and pre-amplified with a $\times$11 gain. Analog to Digital converters with frequency limits of 100-5,000 Hz and a sampling rate of 50 KHz were applied. For a detailed description of the recording setup see \cite{Mahn2011}. Data is transfered to a PC via USB2.0 connection using 16 bit format, in resolution of 27 nV and a rate of 25 KHz and analyzed using Matlab (Mathworks, Natick, MA, USA). 

\subsection{Analyses}
Action potentials were detected on-line by threshold crossing (6 $\times$ standard deviation). The threshold value was defined separately for each of the recording channels at the beginning of an experiment, from a 2 sec long recorded voltage trace. A minimum of 6 msec interval between spikes was required, per electrode. Synchronous events were analyzed off-line by threshold crossing of summed action potentials in 25 msec bins. Threshold value was adjusted relatively to 25\% of active electrodes in the relevant recording area. The initiation time of activity in a given recording region was determined by the average time stamp of the first three spikes in that region. Synfire chains were detected as described by Schrader et al. \cite{Schrader2008}. Briefly, a pair-wise similarity matrix was computed by comparing the set of active electrodes in bins of 3 msec; values were scales between 0-1 by normalizing to the minimal number of active electrodes in each bin. The normalized matrix of similarity was then plotted, color coded such that black depicts full identity between two time bins, and a value of zero is depicted white. In the resulting image, a continuous 45 degree line is the signature of ongoing regularity, while the time gap between such lines represents the period of regularity.
 \vspace{-1em}
\section{acknowledgments} \vspace{-1em}
The authors thank Dani Dagan, Omri Barak, Naama Brener and Noam Ziv for their useful comments and suggestions. This research has received funding from the European Union Seventh Framework Program (FP7/2007-2013) under Grant agreement No. FP7-269459 CORONET.

\newpage
\bibliographystyle{PNASref}  
\bibliography{ref}
\end{document}